\documentclass[preprint,preprintnumbers,amsmath,amssymb,11pt]{revtex4}
\usepackage{graphicx,natbib,color,cancel,soul}
\usepackage{dcolumn}
\usepackage[utf8]{inputenc}
\usepackage{bm,soul}
\usepackage[toc,page]{appendix}

\begin{document}

\title{On the consistency of tachyon warm inflation with viscous pressure}

\author{Antonella Cid}
\email{acidm@ubiobio.cl}
\affiliation{Departamento de F\'{\i}sica, Universidad del B\'io-B\'io, Avenida Collao 1202, Casilla 5-C, Concepci\'{o}n, Chile}

\begin{abstract}
We obtain conditions for the existence of an attractor in the system of equations describing a tachyon warm inflationary model with bulk viscosity taken into account. When these conditions are met the evolution approaches slow-roll regime. We present the primordial power spectrum for the tachyon field by considering a dissipation coefficient depending on the scalar field and temperature.
\end{abstract}

\maketitle

\section{Introduction}

The inflationary paradigm has proved to be successful in solving the problems related with the initial conditions of the standard cosmological model (horizon, flatness, monopoles) \cite{problemsinflation}. Besides it provides a natural explanation to the large scale structures we see today \cite{sdss} and the anisotropy observed in the cosmic microwave background \cite{Planck}. 

In the standard inflationary scenario a single scalar field dominates the dynamics driving a phase of accelerated expansion in the very early universe. During the inflationary period the universe results supercooled and a subsequent phase of reheating is necessary. An alternative scenario is denominated warm inflation \cite{warm}. In this model the dissipative effects are important during inflation and radiation production occurs concurrently with the expansion. The dissipative effect is mediated by a friction term describing the decaying of the scalar field into a thermal bath and the fluctuations of the scalar field arise from thermal rather than quantum contributions \cite{warmfluctuations}. At the end of a period of warm inflation the temperature of the universe is high enough to smoothly go into the radiation dominated phase of the standard cosmological model.

Several models of warm inflation has been studied \cite{warmmodels}, particularly a model of warm inflation driven by a tachyon field was proposed in \cite{0610339}. A tachyon field \cite{tachyon} is an interesting candidate to drive inflationary expansion \cite{tachyoninflation} and it is associated with unstable D-Branes in string theory \cite{string}.

From the theoretical point of view, any inflationary model must be a dynamically plausible solution to the cosmological equations, i.e. a viable inflationary scenario must be an attractor in the phase space of solutions to the cosmological field equations \cite{LiddleParsonsBarrowSalopekBond}.

Usually the slow-roll approximation is used in inflationary models because the equations are considerably simplified. Nevertheless, it is not always verified that the slow-roll solution is an attractor solution to the dynamical set of equations describing the system.

In Refs.\cite{08080261,10070103,12090712,13020168,13115327} the consistency of some warm inflationary models has been revised. Particularly, in Ref.\cite{13115327} the consistency of a tachyon warm inflationary model was considered by assuming a dissipation coefficient independent of temperature. In Ref.\cite{14072604} the primordial spectrum of density fluctuations for a warm tachyon model was calculated taking into account a term of viscous pressure.

Viscous pressure may significantly influence the dynamics of warm inflation  and it is well motivated on physical grounds. A term of viscous pressure may arise from the decay of heavy fields into light fields or by the production of particles by the inflaton \cite {viscous}.

The aim of this work is to show that the slow-roll solution for tachyon warm inflation with bulk viscous pressure is behaved as an attractor under particular conditions and to present the primordial power spectrum of scalar perturbations and the corresponding spectral index for a tachyon warm inflationary model with a dissipation coefficient depending on the temperature.

The paper is organized as follows. In section \ref{basics} we present a model of warm tachyon inflation with viscous pressure. In section \ref{stability} we perform a linear stability analysis, we look for the existence of an attractor in the space of parameters describing the model. In section \ref{fluctuations} we study the associated primordial power spectrum by considering a dissipation coefficient with temperature dependence. In section \ref{example} we present some examples with an exponential potential for the tachyon field. Finally in section \ref{conclusions} we summarize our findings.
\section{Basic Equations}
\label{basics}
We consider a spatially flat, homogeneous and isotropic universe with expansion rate $H$. The matter content is given by a tachyon field $\phi$ and a radiation bath of entropy energy density $s$ and temperature $T$. 
The dynamics  of tachyon warm inflation with viscous pressure is given by \cite{0610339}:
\begin{eqnarray}
\label{H}
3H^2&=&\frac{1}{M_P^2}\left(\frac{V}{\sqrt{1-u^2}}+sT\right)\\
\label{phi}
\frac{\dot{u}}{1-u^2}&=&-\frac{\Gamma}{V}\sqrt{1-u^2}u-3Hu-\frac{V_{,\phi}}{V}\\
\label{rho}
T\dot{s}&=&\Gamma u^2-3H(Ts+\Pi)
\end{eqnarray}
where we have conveniently defined the dimensionless quantity $u=\dot{\phi}$. The evolution is governed by the effective potential $V(\phi,T)$ and a dissipation coefficient $\Gamma(\phi,T)$ ($\Gamma>0$ by the Second Law of Thermodynamics). We consider the bulk viscous pressure given by the relation $\Pi=-3\zeta H$, where the phenomenological coefficient of bulk viscosity $\zeta$ is assumed as a function of the energy density $\rho$ and it is positive defined. The subscript in a function denotes derivative, we work in natural units and we consider $8\pi G=M_P^{-2}$.  In Eqs.(\ref{H}-\ref{rho}), the tachyon field has dimension $[\phi]=M^{-1}$, the dissipation coefficient $[\Gamma]=M^5$, and the bulk viscous pressure $[\Pi]=M^4$, where $M$ denotes the dimension of mass.

The energy density of the tachyon field is defined as $\rho_\phi=\frac{V}{\sqrt{1-u^2}}$. We note that the weak energy condition for this fluid impose $u^2<1$. On the other hand, the energy density of the radiation bath is related to the entropy density by $\gamma\rho=sT$ where $\gamma$ corresponds to the adiabatic coefficient, which must slowly vary with expansion but for the sake of simplicity, we will consider it constant and in the range $1<\gamma<2$.

From the thermodynamic relation $s=-\frac{\partial f}{\partial T}$, where $f$ is the Helmholtz free energy, we can relate the entropy density with the effective potential as $s\approx -\frac{\gamma}{4(\gamma-1)}V_T$ \cite{10070103}. 

Basically, the slow-roll approximation consists of consider a dominant potential and neglect terms of the highest order in time derivatives from Eqs.(\ref{H})-(\ref{rho}). The validity of this approximation will depend on a set of dimensionless slow-roll parameters defined in terms of the potential as:
\begin{eqnarray}
\epsilon=\frac{M_P^2}{2}\frac{V_{,\phi}^2}{V^3},\ \ \ \ \ \eta=M_P^2\frac{V_{,\phi\phi}}{V^2},\ \ \ \ \ \beta=M_P^2\frac{V_{,\phi}\Gamma_{,\phi}}{V^2\Gamma}
\end{eqnarray}

In order to give account of the temperature dependence in the effective potential and in the dissipation coefficient it is necessary to introduce two additional dimensionless parameters \cite{0305015}:
\begin{eqnarray}
b=\frac{V_{,\phi T}T}{V_{,\phi}},\ \ \ \ \ c=\frac{\Gamma_{,T}T}{\Gamma}
\end{eqnarray}

During the inflationary period the Hubble expansion rate is nearly constant $\vert\frac{\dot{H}}{H^2}\vert\ll1$, this defines the first slow-roll parameter $\epsilon_H$ which we can write from Eqs.(\ref{H})-(\ref{rho}) as:
\begin{eqnarray}
\epsilon_H=-\frac{\dot{H}}{H^2}=\frac{3u^2+\left(\frac{sT}{V}+\frac{\Pi}{V}\right)\left(4\sqrt{1-u^2}-1\right)-\frac{\Gamma u^2}{3HV}\left(1-\sqrt{1-u^2}\right)}{2\left(1+\frac{sT}{V}\sqrt{1-u^2}\right)}
\end{eqnarray}
 The second slow-roll parameter is given by:
\begin{eqnarray*}
\eta_H&=&-\frac{\ddot{H}}{\dot{H}H}=\frac{\dot{V}}{HV}+\frac{\dot{u}}{Hu}\frac{u^2}{1-u^2}\left(1-
\frac{\sqrt{1-u^2}\left(\frac{\Gamma u^2}{3HV}-\frac{sT}{V}-\frac{\Pi}{V}\right)}
{3u^2+\left(\frac{sT}{V}+\frac{\Pi}{V}\right)\left(4\sqrt{1-u^2}-1\right)-\frac{\Gamma u^2}{3HV}\left(1+\sqrt{1-u^2}\right)}\right)\\
&&-\frac{\left(2u^2(-3+\frac{\Gamma}{3HV}(\sqrt{1-u^2}-1))\frac{\dot{u}}{uH}+\frac{\left(\Gamma/3HV\right)^{\cdot}}{H}u^2(\sqrt{1-u^2}-1)-\frac{\left(sT/V+\Pi/V\right)^{\cdot}}{H}(4\sqrt{1-u^2}-1)\right)}
{3u^2+\left(\frac{sT}{V}+\frac{\Pi}{V}\right)\left(4\sqrt{1-u^2}-1\right)-\frac{\Gamma u^2}{3HV}\left(1-\sqrt{1-u^2}\right)}
\end{eqnarray*}

The smallness of these parameters give the relative size of terms neglected in the slow-roll approximation. We note that in order to have $\epsilon_H\ll1$ (or accelerated expansion) we must impose $u^2\ll1$, $sT\ll V$, $\Pi\ll V$ and $\Gamma u^2\ll3HV$; furthermore the condition $\eta_H\ll1$ implies $\dot{u}\ll Hu$, $\dot{\Gamma}\ll H \Gamma$ and $\dot{\Pi}\ll H\Pi$. Additionally we impose that the decay of tachyon field into radiation is quasi-stable, $\dot{s}\ll sH$.
The end of inflation is presented when $\epsilon_H$ becomes 1 or accelerated expansion ends.

In the slow-roll regime Eqs.(\ref{H})-(\ref{rho}) reduce to:
\begin{eqnarray}
&&3H^2=\frac{V}{M_P^2}\label{sr1}\\
&&3H(1+r)u+\frac{V_{,\phi}}{V}=0\label{phisr}\label{sr2}\\
&&3H(sT+\Pi)-\Gamma u^2=0\label{sTsr}
\label{sr3}
\end{eqnarray}
where $r$ is a dimensionless coefficient defined as $r=\frac{\Gamma}{3HV}$. This coefficient allow us to distinguish between two regimes of warm inflation, the weak dissipation regime ($r\ll1$) and the strong dissipation regime ($r\gg1$). We will focus on this latter case because in the weak regime viscosity contributions are unimportant \cite{viscous}.

From Eq.(\ref{sTsr}) we note that a consistency condition for $\Pi<0$ is that $\vert\frac{\Pi}{sT}\vert\le1$.

It is worth to mention that in the slow-roll and strong dissipation regime Eqs.(\ref{sr1})-(\ref{sr3}) look alike the slow-roll equations in the case of warm inflation with a canonical scalar field.

The number of e-foldings is given by:
\begin{eqnarray}
N(\phi)=-\frac{1}{M_P^2}\int_{\phi}^{\phi_e}\frac{V^2}{V_{,\phi}}(1+r)d\phi
\end{eqnarray}
where the subscript $e$ denotes the end of inflation.
\section{Stability Analysis}
\label{stability}
Following Ref.\cite{12090712}, we perturb Eqs.(\ref{phi}) and (\ref{rho}) around the slow roll solution $x_0$ and express the system in matrix form as, 
\begin{eqnarray}
\label{pert}
\delta X'=M(x_0)\delta X-x'_0
\end{eqnarray}
where for convenience we have considered derivatives with respect to $\phi$ denoted with primes and 
\begin{eqnarray}
M(x_0)= \left( \begin{array}{cc}
A & B  \\
C & D  \end{array} \right)_{x_0},
\ \ \ \ 
\delta X = \left( \begin{array}{c}
\delta u   \\
\delta s   \end{array}\right) 
\ \ \ \ \textrm{and} \ \ \ \ 
x'_0 = \left( \begin{array}{c}
u'_0   \\
s'_0   \end{array}\right) 
\label{matrix}
\end{eqnarray}
The term $x_0'$ is present because the slow-roll solution is not an exact solution to Eqs.(\ref{H})-(\ref{rho}).

In order that the slow-roll solution becomes an attractor to the system, the eigenvalues of $M(x_0)$ must be negative and the components of vector $x_0'$ must be small. The condition on the eigenvalues is met when: the determinant of matrix $M(x_0)$ is positive and the trace is negative.

Given that the slow-roll solution is not an exact solution to Eqs.(\ref{H})-(\ref{rho}) the term $x_0'$ in (\ref{pert}) must be small. The size of $x_0'$ depends on the quantities, 
\begin{eqnarray}
\frac{\dot{s}}{Hs}&=&
\frac{3}{\Delta}\left(\left(\frac{2r+8}{r+1}-\frac{1}{\chi} \right) \epsilon -2\eta +\frac{r-1}{r+1} \beta +\frac{3}{4}b\chi_l \frac{(1+r)^2}{r}-\frac{3\gamma}{4}bc\chi\frac{(1-r^2)}{r}\right)\\
\frac{\dot{u}}{Hu}&=&
\frac{3}{\Delta}\left(\frac{\epsilon \left(3\frac{\chi_l}{\chi}-c\omega(3 +r (-\frac{1}{\chi}+2))\right)}{(1+r)}+\left(c\omega-\frac{\chi_l}{\chi}\right) \eta+\frac{ r \beta\chi _l}{(1+r)\chi}+\frac{3}{4} bc (1+r)\chi _l \right)
\end{eqnarray}
where we define $\sigma=\frac{\Pi}{sT}$, $\chi=1+\sigma$,  $\chi_l=\gamma(1+l\sigma)$, $\omega=\gamma-1$ and $\Delta=3\gamma(r+1)\chi_l-3\omega(c(1-r)+2 b(r+1))+\frac{4\epsilon}{1+r}\frac{r\omega^2}{\gamma\chi^2}\left(1+\frac{8+2r}{1+r}\chi\right)$. We note that the conditions $\vert\frac{\dot{s}}{Hs}\vert\ll1$ and $\vert\frac{\dot{u}}{Hu}\vert\ll1$ are satisfied in the strong dissipation regime provided that:
\begin{eqnarray}
\label{R2}
\vert\epsilon\vert\ll r, \ \ \vert\eta\vert\ll r,\ \ \vert\beta\vert\ll r,\ \ \vert b\vert\ll1
\end{eqnarray}

For tachyon warm inflation with viscous pressure we have:
\begin{eqnarray}
A&=&\frac{H}{u}\left(-3(1+r)+\frac{\epsilon(4r+1)}{(1+r)^2}\right)\\
B&=&\frac{H}{s}\left(-3(\gamma-1)\left(cr-b(1+r)\right)+\frac{3}{2}\alpha\sigma\tilde{\sigma}+\frac{r\epsilon(\gamma-1)}{(1+r)^2}\left(\frac{\alpha(1-\sigma)}{\gamma-1}+3c+\frac{4(\gamma-1)}{\gamma}\right)\right)\\
C&=&\frac{Hs}{u^2}\left(6+6\sigma-\frac{\epsilon(1+2\sigma)}{(1+r)^2}\right)\\
D&=&\frac{H}{u}\left(3(\gamma-1)(c-1+\sigma c)-3(1+\gamma\sigma l)-\frac{3}{2}\beta\sigma\tilde{\sigma}+\frac{\epsilon(1+\sigma)}{(1+r)^2}\left(r\beta-\frac{2\sigma\tilde{\sigma}(\gamma-1)^2}{\gamma(1+\sigma)}\right)\right)
\end{eqnarray}
where we define $\alpha=\frac{12+\gamma(11\gamma-24)}{\gamma}$, $\beta=\frac{(2-3\gamma)(2-\gamma)}{\gamma}$ and $\tilde{\sigma}=\frac{\Pi}{V}=\frac{2r\epsilon\sigma}{3(1+\sigma)(1+r)^2}$ and we have considered $\zeta(\rho)\propto \rho^{l}$.

The requirement of positive determinant of matrix $M$ in (\ref{matrix}) is more restrictive than the constraint over the trace, we find that in order to the slow-roll regime behaves as an attractor in the strong dissipation regime ($r\gg1$) we must impose the following condition:
\begin{eqnarray}
\label{R1}
\vert c\vert <\frac{2\gamma(1+\sigma l)-\beta\sigma\tilde{\sigma}-4b(1+\sigma)(\gamma-1)}{2(1+\sigma)(\gamma-1)}
\end{eqnarray}
where we have considered $b>0$, $\vert \sigma\vert <1$ and $\vert \sigma l\vert <1$. 

We note that the parameter $c$ is not required to be small and the condition on the parameter $b$ is more restrictive than the conditions over the other parameters. It is worth to mention that the condition on $b$ restricts the temperature corrections of the potential to be small.

We note that the result for a canonical scalar field \cite{08080261} is recovered in the limit $\sigma\rightarrow0$, and the result in Ref.\cite{10070103} for a canonical scalar field with viscous pressure is reproduced by taking $l\rightarrow0$.

\section{Density Fluctuations}
\label{fluctuations}
The authors of Ref.\cite{14072604}  have determined the primordial power spectrum of density fluctuations for the case of a tachyon field with a term of bulk viscous pressure:
\begin{eqnarray}
\label{spectrum}
\delta_H^2=\left(\frac{2}{15}\right)^2\frac{M_P^4e^{-2F(\phi)}}{H ^2(1+r)^2\dot{\phi}^2}\langle\delta\phi\rangle^2
\end{eqnarray}
where $\langle\delta\phi\rangle^2$ are the corresponding fluctuations of the tachyon field and 
\begin{eqnarray}
&&F(\phi)=-\int \left[A(\phi)+B(\phi)\right]d\phi, \ \ \textrm{where}\ \ A(\phi)=\frac{\left(\Gamma/V\right)_{\phi}}{3H+\Gamma/V},\\
&&B(\phi)=\frac{9}{8G(\phi)}\frac{2H+\Gamma/V}{\left(3H+\Gamma/V\right)^2}\left(\Gamma+4H V-\left(\gamma-1+\Pi\frac{\zeta_{\rho}}{\zeta}\right)\frac{\Gamma_{\phi}\left(\ln V\right)_{\phi}}{3\gamma H\left(3H+\Gamma/V\right)}\right)\frac{\left(\ln V\right)_{\phi}}{V}\\
&&G(\phi)=1-\frac{2sT+3\Pi+\frac{1}{\gamma}\left(sT+\Pi\right)\left(\Pi\frac{\zeta_{\rho}}{\zeta}-1\right)}{8H^2}
\end{eqnarray}
The fluctuations of the tachyon field are generated by thermal interaction with the radiation field, rather than quantum fluctuations. In the case of a dissipation coefficient depending on the temperature it is hard to get analytical solutions. The authors of Ref.\cite{09053500} give an approximate expression for their numerical result of the power spectrum for a canonical scalar field. These results are derived to leading order in the slow-roll approximation. It is possible to develop an analogous procedure in the case of tachyon warm inflation with viscous pressure. To leading order in slow-roll the perturbation equations for the tachyon field and the radiation bath can be simplified as it is shown in the Appendix. The contribution of the viscous pressure is parametrized with the $\sigma$ parameter defined in section \ref{stability}, which we assume approximately constant during inflation.
\begin{eqnarray}
\label{thermal}
\langle\delta\phi\rangle^2=\frac{\sqrt{3\pi}}{2}HT\sqrt{1+r}\left(1+\frac{r}{r_\delta}\right)^{\delta}
\end{eqnarray}
where $\delta=\frac{3\gamma c}{4(\omega+\gamma l\sigma)}$ and $r_{\delta}$ is obtained numerically (see Appendix).
\begin{table}[ht!]
\begin{tabular}{|c|c|c|c|}
\hline
$r_{\delta}$&$l=0$	& $l=0.5$	& $l=1$\\ \hline
$c=1$		&8.53	& $13.81$	& 41.70\\ \hline
$c=2$		&7.66	& $14.78$	& 33.93\\ \hline
$c=3$		&7.27	& $15.85$	& 25.88\\ \hline
\end{tabular}
\caption{\label{tablei} Numerical values of $r_{\delta}$, we assume $\gamma=4/3$ and $\sigma=-0.1$}
\end{table}

The Eq.(\ref{thermal}) is consistent with the spectrum of the scalar field with a dissipation coefficient independent of the temperature \cite{warmfluctuations}. We note that for $\Pi=-3\zeta_0H$ the result in Ref.\cite{09053500} is valid in the case of viscous pressure. The result in Eq.(\ref{thermal}) is valid for $\delta>0$ or $\sigma>-\frac{\omega}{\gamma l}$.

From Eqs.(\ref{spectrum}) and (\ref{thermal}) we calculate the new spectral index in the strong dissipation regime,
\begin{eqnarray}
n_s-1=\frac{d\ln \delta_H^2}{d\ln k}=b p_1+\tilde{\epsilon} p_2+\tilde{\eta} p_3+\tilde{\beta} p_4-2\left(\tilde{\beta}-2\tilde{\epsilon}+\frac{9}{4}\tilde{\epsilon}\right)
\label{spectral}
\end{eqnarray}
where we define $\tilde{\beta}=\frac{\beta}{r}$, $\tilde{\epsilon}=\frac{\epsilon}{r}$, $\tilde{\eta}=\frac{\eta}{r}-\tilde{\epsilon}$ and
\begin{eqnarray}
p_1&=&-\frac{3 \gamma  (1+\sigma) (1+l \sigma ) \left(4\omega+2 c \omega+3 c^2 \gamma +2 (2+c) l \gamma  \sigma \right)}{16 (\omega +l \gamma  \sigma ) (\gamma +l \gamma  \sigma +c\omega (1+\sigma ))}\\
p_2&=&\frac{3 c^2\omega \gamma  (2+3 \sigma )
-2 (\omega +l \gamma  \sigma ) (5 \gamma -2+3 l \gamma  \sigma )
-c \left(8 \omega^2-9 \gamma ^2+\sigma  \left(6 \omega^2-l \gamma  (8+\gamma )+6 l\omega \gamma  \sigma \right)\right)}{4 (\omega +l \gamma  \sigma ) (\gamma +l \gamma  \sigma +c\omega (1+\sigma ))}\\
p_3&=&-\frac{3 c^2\omega \gamma  (1+\sigma )-2 c \omega (1+\sigma ) (\omega +l \gamma  \sigma )-4 (\omega+l \gamma  \sigma ) (1+(1+(-1+l) \gamma ) \sigma )}{2 (\omega+l \gamma  \sigma ) (\gamma +l \gamma  \sigma +c\omega	 (1+\sigma ))}\\
p_4&=&-\frac{3 c \gamma ^2 (1+l \sigma )+2 (\omega+l \gamma  \sigma ) (2-\gamma +(2+(-2+l) \gamma ) \sigma )}{4 (\omega+l \gamma  \sigma ) (\gamma +l \gamma  \sigma +c \omega(1+\sigma ))}
\end{eqnarray}
We have considered that the interval of wave number is related with the number of e-foldings $N$ by the relation $d\ln k=dN$.
\section{Tensor Perturbations}
Tensor perturbations do not couple to the thermal background, so gravitational waves are only generated by quantum fluctuations, as in standard fluctuations \cite{0006077},
\begin{eqnarray}
\mathcal{P}_T=\frac{2}{M_P^2}\left(\frac{H}{2\pi}\right)^2=\frac{V}{6\pi^2 M_P^4}
\end{eqnarray}
The tensor spectral index is given as usual by $n_T=-2\tilde{\epsilon}$.

The tensor-scalar ratio for tachyon warm inflation in the strong dissipation regime is given by,
\begin{eqnarray}
R=\frac{\mathcal{P}_T}{\mathcal{P}_R}=\frac{2\sqrt{3}}{\pi^2\sqrt{\pi}}\frac{\tilde{\epsilon}H^3\sqrt{r}}{T}e^{2F}\left(\frac{r}{r_{\delta}}\right)^{\delta}
\end{eqnarray}
where $\mathcal{P}_R=\frac{25}{4}\delta_H^2$.

\section{Example}
\label{example}
In order to maintain the validity of the slow-roll approximation as an attractor in the strong dissipation regime, the condition on the slow-roll parameter $b$ guarantees that the thermal corrections to the effective potential must be small. This allow us to consider separable potential as $V(\phi,T)=V_1(T)+V_2(\phi)$. On the other hand, by using Noether symmetry the authors of Ref.\cite{08092331} found that if a tachyon field is driving accelerated expansion, then the potential associated to the field must be exponential. The above arguments motivate the use of the following potential:
\begin{eqnarray}
V(\phi)=V_0 e^{-\alpha \phi}, 
\end{eqnarray}
where $V_0$ and $\alpha$ are positive constants and for the sake of simplicity we have only considered a potential depending on the scalar field.

Taking into account the results of section \ref{fluctuations} we consider a dissipation coefficient of the form:
\begin{eqnarray}
\label{GammaAnz}
\Gamma(\phi,T)=\Gamma_0e^{-\frac{3}{2}\alpha\phi} T^c
\end{eqnarray}
where $\Gamma_0$ and $c$ are constants positive defined. 

In this exmaple we also consider $\Pi=-\zeta_0\rho H$.

Usually, the end of a period of slow-roll inflation is implemented through the condition $\epsilon_H=1$ or equivalently by imposing a null acceleration to the system. It is not always possible to find this behaviour because $\epsilon_H$ must be an increasing function in order to reach the condition at the end of a long enough period of inflation \cite{13115327}. In the case of a bulk viscosity coefficient proportional to the radiation density, it is possible to analytically show that a dissipation coefficient depending on the temperature allows to achieve this behaviour. From Eqs.(\ref{sr1})-(\ref{sr3}) and Eq.(\ref{GammaAnz}) we have:
\begin{eqnarray}
\epsilon_H=-\frac{\dot{H}}{H^2}=\frac{V_0 \alpha ^2	e^{\frac{1}{2} \alpha  \phi} }{2 \Gamma_0}\left(\frac{C_r \Gamma_0}{M_P^2 V_0^{3/2}\alpha^2} \left(\sqrt{3} M_P \gamma -3 e^{-\frac{1}{2} \alpha  \phi} \sqrt{V_0} \zeta_0\right)\right)^{\frac{c}{4+c}}
\end{eqnarray}
where we assumed that the energy density is related to the temperature through the relation $\rho=C_r T^4$. Given that $\phi$ is an increasing function, $\epsilon_H$ is an increasing function provided that $\sigma=-\frac{3\zeta_0H}{\gamma}<1$, i.e. $\epsilon_H<1$ during inflation and one at the end of inflation.

From Eqs.(\ref{sr1})-(\ref{sr3}) and considering $\rho=C_r T^4$ we can explicitly find:
\begin{eqnarray}
T(H)&=&\left(\frac{\sqrt{3}C_r\Gamma_0}{M_PV_0^{3/2}\alpha^2}(\gamma-3\zeta_0H)\right)^{-\frac{1}{4+c}}\\
r(H)&=&\frac{M_P\Gamma_0}{\sqrt{3}V_0^{3/2}}\left(\frac{\sqrt{3}C_r\Gamma_0}{M_PV_0^{3/2}\alpha^2}(\gamma-3\zeta_0H)\right)^{-\frac{c}{4+c}}
\end{eqnarray}
We note that during inflation the temperature is a decreasing function, so is $\rho$.

Given that in our example we consider the temperature dependence in $\Gamma$, it is not possible to find more explicit results. Several examples where $\Gamma$ depends only on $\phi$ were studied in Ref.\cite{14072604}. The current constraints on the amplitude of the primordial power spectrum for scalar perturbations $\mathcal{P}_R(k_0)=2.23\times10^{-9}$ and the tensor to scalar ratio $R(k_0)<0.11$ \cite{13035076} (for a pivot scale $k_0=0.002$Mpc$^{-1}$), allow us to constrain the value of the potential at the Hubble horizon crossing to be $V_*<10^{-8}M_p^4$.

In our example the slow-roll parameters are given by:
\begin{eqnarray}
\tilde{\epsilon}=\tilde{\eta}=\frac{\tilde{\beta}}{3}=\frac{M_p^2\alpha^2}{2rV}\ll1, \ \ b=0
\end{eqnarray}
and we can rewrite the spectral index in (\ref{spectral}) as:
\begin{eqnarray}
n_s-1=\frac{M_P^2\alpha^2}{2rV}\left(p_2-\frac{1}{2}+p_3+3(p_4-2)\right)
\end{eqnarray}
We note that $n_s-1$ is always negative for $l=1$, $\gamma=4/3$ and $c=1,2,3$, $-0.25<\sigma<0$. Consequently in a scenario of tachyon warm inflation with viscous pressure it is common to have a red tilted primordial power spectrum for scalar perturbations.

\section{Final Remarks}
We analysed a model of tachyon warm inflation by taking into account a term of viscous pressure. In the strong dissipation regime we found conditions in order to the slow-roll solution behaves as an attractor to the dynamical system of equations describing the model. The slow-roll regime is implemented by imposing restrictions on the size of the  five slow-roll parameters of the model associated to the effective potential and the dissipation coefficient. The conditions on $\epsilon$, $\eta$, $\beta$ and $b$ are the standard ones for warm inflation \cite{08080261}, but the condition on the $c$ parameter has contributions associated to the viscous pressure term and it is less restrictive than the case without viscous pressure \cite{13115327} for $-1<\sigma<0$ and $l<1$.

We calculated an approximated expression for the primordial power spectrum of a tachyon field driving a period of warm inflation. We consider a temperature dependence in the dissipation coefficient and viscous pressure. This result follow from Ref.\cite{09053500} and the numerical integration of the corresponding perturbed equations is expected to be performed in future work.

We presented the primordial power spectrum for density fluctuations by considering a dissipation coefficient with temperature dependence, and we calculate the corresponding spectral index to this model. We work through a specific example with a decaying exponential potential for the tachyon field. We found that it is possible to build up a consistent model with temperature dependence in the dissipation coefficient, which properly finish the inflationary period.
\label{conclusions}

\begin{acknowledgments}
The author dedicates this article to the memory of Dr. Sergio del Campo. This work has been partially supported by Fondecyt grant N$^\circ$ 11110507 and Universidad del Bio-Bio through grant DIUBB 121407 GI/VC.
\end{acknowledgments}

\begin{appendices}
For tachyon warm inflation with viscous pressure Eqs.(45) and (46) of Ref.\cite{09053500} become:
\begin{eqnarray*}
\delta\phi''-(3r+2)\frac{\delta\phi'}{z}+\delta\phi+3c\frac{3\gamma}{4}(\Gamma\dot{\phi})^{-1}\frac{\delta\rho_r}{z^2}=\sqrt{2\Gamma_{eff}T}\hat{\xi}\\
\delta\rho_r''-(3\gamma(1+l\sigma-c/4)+4)\frac{\delta\rho_r'}{z}+(15\gamma(1+l\sigma-c/4))\frac{\delta\rho_r}{z^2}+(\omega+\gamma l\sigma)\delta\rho_r+(\Gamma\dot{\phi})\delta\phi=0
\end{eqnarray*}
In calculating the primordial spectrum of the tachyon field at horizon crossing, we get an expression analogous to Eq.(59) of Ref.\cite{09053500}:
\begin{eqnarray*}
P_{\phi}^{c\neq0}(k,1)\approx\left(
\frac{G^{31}_{13}
\left(W\middle|
\begin{matrix}1-\frac{8\delta}{9\gamma^2} & &\\ Z & 0 &\frac{5}{2} \end{matrix}
\right)
}{\left(3W^{-\frac{16}{9\gamma^2}}\right)^{-\delta}
\Gamma_R\left(\frac{8\delta}{9\gamma^2}+Z\right)\Gamma_R\left(\frac{8\delta}{9\gamma^2}\right)\Gamma_R\left(\frac{8\delta}{9\gamma^2}+\frac{5}{2}\right)
}\right)^2\frac{\Gamma_R\left(\delta+\frac{3}{2}\right)}{\Gamma_R\left(\frac{3}{2}\right)}r^{\delta}P_{\phi}^{c=0}(k,1)
\end{eqnarray*}
where $\delta=\frac{3\gamma c}{4(\omega+\gamma l \sigma)}$, $W=\frac{\omega+\gamma l\sigma}{4}$, $Z=\frac{3\gamma}{2}\left(1+l\sigma-\frac{c}{4}\right)$, $\Gamma_R$ is the Gamma function and $G^{31}_{13}$ is a Meijer-G function. From this last expression we obtain the primordial power spectrum of the tachyon field. Some numerical values are shown in TABLE \ref{tablei}.
\end{appendices}

\end{document}